\documentstyle[12pt]{article}

\author{M.I.Krivoruchenko\\
Institute for Theoretical and Experimental Physics,\\
B.Cheremushkinskaya, 25, 117259 Moscow, Russia\\
(e-mail: mikhail@vxitep.itep.ru)}
\title{Suppression of the $n-\bar{n}$ Oscillation in Nuclei
}
\date{}
\begin{document}

\maketitle
\begin{abstract}
The problem of nuclear decays occurring due to the neutron-antineutron
oscillation is analyzed in the optical potential model and within the
field-theoretic S-matrix approach. The result of the optical potential model
for the nuclear decay width is rederived within the field-theoretic S-matrix
approach. The n-\=n oscillation in nuclei is suppressed drastically. We
discuss an example for the spin precession of an atom, in which analogous
suppression takes place due to transparent physical reasons.
\end{abstract}

\vspace{5mm}


\newpage

\setcounter{equation}{0}

The possible occurrence of the baryon-number-violating processes predicted
by the grand unified theories (GUT's) stimulated searches for the
neutron-antineutron oscillation$^{1-8}$. The limits for the neutron
oscillation time in the vacuum, $\tau =1/\epsilon >1yr$, are derived from
stability of nuclei$^{1-9}$ and experiments with free neutrons$^{10}$.

The n-\=n oscillation is described by the Hamiltonian

\begin{equation}
\label{1}V=\frac 12\epsilon \int d{\bf x}\bar \Psi ({\bf x)}\sigma _1\Psi (%
{\bf x)}\,
\end{equation}
where $\sigma _1$ is the Pauli matrix and

\begin{equation}
\label{2}\Psi ({\bf x})=\left(
\begin{array}{c}
\psi _n(
{\bf x}) \\ \psi _{\overline{n}}({\bf x})
\end{array}
\right)
\end{equation}
is a doublet of the neutron and antineutron fields. The $V$ is Hermitian,
since $\psi _{\overline{n}}({\bf x})=(\psi _n({\bf x}))_c=\psi _n({\bf x}%
)^{+}$(in the Majorana representation).

For nuclei with atomic numbers $A\ >>\ 1$, the decay width due to the n-\=n
oscillation takes the form$^{1-8}$

\begin{equation}
\label{3}\Gamma =\epsilon ^2N\frac{\Gamma _A}{(M-\bar M)^2+\Gamma _A^2/4}.\,
\end{equation}
Here $N\ =A-Z$ is the number of neutrons in the nucleus, $Z$ the number of
protons, $M$ the nuclear mass, and $\bar M$ the average mass of the state
obtained from the nucleus by the replacement of a neutron by an antineutron.
The value $\Gamma _A$ is a hadron-scale antineutron annihilation width of
the intermediate state. The nuclear decay rate per neutron, $\Gamma
/N=\epsilon \xi $, contains an additional small parameter

\begin{equation}
\label{4}\xi =\frac{\epsilon \Gamma _A}{(M-\bar M)^2+\Gamma _A^2/4}%
<10^{-31}\,
\end{equation}
as compared to the rate $\epsilon $ of the vacuum n-\=n oscillation.

In recent paper$^{11}$ the finite-time S-matrix approach is developed for
evaluation of the nuclear decay rate due to the neutron oscillation. It is
claimed that the optical potential model is in disagreement with the
field-theoretic S-matrix approach. We analyze more carefully the optical
potential model, derive the Eq.(3) within the infinite- and finite-time
S-matrix approaches, and clarify physical nature of the mechanism
responsible for occurrence of the small parameter $\xi $. The potential
model is shown to be in agreement with the S-matrix theory.

We shall be interested in solutions to the evolution problem with the
Hamiltonian of the form

\begin{equation}
\label{5}H=\epsilon \sigma _1+U_nP_{+}+U_{\bar n}P_{-}-i\Gamma (t)/2P_{-}\,
\end{equation}
where $P_{\pm }=(1\pm \sigma _3)/2$ are projection operators to the neutron
and antineutron states, $\Gamma (t)$ is a time dependent width. The real
potentials, $U_n$ and $U_{\bar n}$, describe the neutron and antineutron
rescattering effects on the surrounding nucleons. The potential model for
the in-medium neutron oscillation is described by the Hamiltonian (5) with
constant width $\Gamma (t)=\Gamma _A$. The time dependent width $\Gamma (t)$
is considered for further applications (see below).

The Hamiltonian (5) can be split into two parts, $H=H_0+V$ with $H_0$ = $%
U_nP_{+}+U_{\bar n}P_{-}-i\Gamma (t)/2P_{-}$ and $V=\epsilon \sigma _1$. In
the interaction representation, the potential $V$ takes the form

\begin{equation}
\label{6}V_I(t)=U^{-1}(t,0)VU(t,0)\,
\end{equation}
with $U(t,0)=\exp (-i\int H_0(t^{\prime })dt^{\prime
})=e^{-iU_nt}P_{+}+e^{-iU_{\bar n}t-\gamma (t)}P_{-}$ and $\gamma (t)=\frac
12\int \Gamma (t^{\prime })dt^{\prime }$. The potential $V_I(t)$ can be
found to be

\begin{equation}
\label{7}V_I(t)=\epsilon (e^{-i\Delta Ut-\gamma (t)}\sigma _{+}+e^{+i\Delta
Ut+\gamma (t)}\sigma _{-})\,
\end{equation}
where $\Delta U=U_{\bar n}-U_n=\bar M-M$ and $\sigma _{\pm }=(\sigma _1\pm
i\sigma _2)/2$.

The transition amplitudes are determined by the S-matrix elements in the
Heisenberg representation: $S(t,0)=U(t,0)T\exp (-i\int V_I(t^{\prime
})dt^{\prime })$. To the second order in the $\epsilon $, we get

\begin{equation}
\label{8}
\begin{array}{c}
S(t,0)=e^{-iU_nt}(P_{+}+e^{-i\Delta Ut-\gamma (t)}P_{-}-i\epsilon
(F_{-}(t)\sigma _{+}+e^{-i\Delta Ut-\gamma (t)}F_{+}(t)\sigma _{-}) \\
-\epsilon ^2(G_{-}(t)P_{+}+e^{-i\Delta Ut-\gamma (t)}G_{+}(t)P_{-})+...)
\end{array}
\end{equation}
where

\begin{equation}
\label{9}
\begin{array}{c}
F_{\pm }(t)=\int_0^te^{\pm i\Delta Ut^{\prime }\pm \gamma (t^{\prime
})}dt^{\prime }, \\
G_{\pm }(t)=\int_0^te^{\pm i\Delta Ut^{\prime }\pm \gamma (t^{\prime
})}F_{\mp }(t)dt^{\prime }.
\end{array}
\end{equation}

The disappearance of the neutrons is described by a projection $S(t,0)P_{+}$
of the S-matrix to the neutron component of the initial wavefunction (2).
Using Eqs.(8) and (9), we obtain for $\Gamma (t)=\Gamma _A$ and $t>>1/\Gamma
_A$

\begin{equation}
\label{10}S(t,0)P_{+}=e^{-iU_nt}(P_{+}-i\epsilon /\chi \sigma _{-}-\epsilon
^2t/\chi P_{-}+...)\,
\end{equation}
where $\chi =\Gamma _A/2+i(\bar M-M)$. The decay probability is determined
by absolute square of the S-matrix element $|<\psi _n|S(t,0)|\psi
_n>|^2=1-\epsilon \xi t+...$, where the use is made of the relation $%
\epsilon Re\{2/\chi \}=\xi $. Comparison with the decay law for
quasistationary states, $|<Q|S(t,0)|Q>|^2=\exp (-\Gamma _Qt)=1-\Gamma
_Qt+... $ at $t<<1/\Gamma _Q$ allows to fix the nuclear decay width per
neutron, $\Gamma /N=\epsilon \xi $. In this way we reproduce Eq.(3).

The potential model operating with the non-Hermitian Hamiltonian (5) gives
the phenomenological description for the nuclear decays. The results of the
potential model can be justified within the field-theoretic S-matrix
approach.

Let $|N,Z$$>$ be a discrete energy eigenstate of the strong Hamiltonian $H_0$%
, associated to a nucleus of mass $M,$ charge $Z$, atomic number $A=N+Z$,
and zero momentum. With respect to the total Hamiltonian $H=H_0+V$ (in the
field-theoretic S-matrix approach, the $H_0$ and $V$ are Hermitian), the
state $|N,Z$$>$ is no longer an energy eigenstate. In what follows, the
ceter-of-mass variables are factored out, so the normalization condition
reads simply $<N,Z|N,Z>=1$. The decay probability of the state $|N,Z>$ to a
channel $X$ in the first order to the $V$ is determined by the matrix element

\begin{equation}
\label{11}<X|V|N,Z>=\epsilon \sqrt{N}g_XA(M_X).\,
\end{equation}
The state $|X>$ consists of $N-2$ neutrons, $Z$ protons, and a number of
mesons. It belongs to the continuum energy spectrum of the $H_0$. The factor
$\sqrt{N}$ takes into account structure of the operator $V$. The function $%
A(M_X)$ has the Lorentz form (cf. Ref.12, Ch.27)

\begin{equation}
\label{12}A^2(M_X)=\frac{\Gamma _A}{(M-\bar M)^2+\Gamma _A^2/4}.\,
\end{equation}
The state $V$$|N,Z>$ has indefinite energy and definite (zero) momentum,
since $[H_0,V]\not =0$ and $[{\bf P},V]=0$. The value $\bar M$ is the
average mass of this state. The scattering states $|X>$ are grouped around
the invariant mass $M_X\approx \bar M$. The function $A(M_X)$ reflects
spread in the $M_X$ of the scattering states $X$ entering a superposition to
form the localized wave packet $V|N,Z>$. The state $V|N,Z>$ decays with a
strong width $\Gamma _A$. The value $g_X$ is a coupling constant to the
channel $X$. This is a smooth function of the $M_X$.

In the nonrelativistic approximation, the state $V$$|N,Z>$ consists of the
fixed number of particles. Its normalization condition reads $%
<N,Z|V^2|N,Z>=\epsilon ^2N$. Inserting in this equation complete set of the
states $|X>$, we obtain

\begin{equation}
\label{13}\int \frac{dM^{\prime }}{2\pi }A^2(M^{\prime })e(M^{\prime })=1\,
\end{equation}
where

\begin{equation}
\label{14}e(M^{\prime })=\sum_X\int g_X^2(2\pi )^4\delta ^4(P_X-P^{\prime
})d\tau _X,\,
\end{equation}
$P^{\prime }=(M^{\prime },{\bf 0})$, and $d\tau _X$ is element of the phase
space.

The expression for the nuclear width,

\begin{equation}
\label{15}\Gamma =\sum_X\int |<X|V|N,Z>|^2(2\pi )^4\delta ^4(P_X-P)d\tau
_X\,
\end{equation}
where $P=(M,{\bf 0})$ with the use of Eqs.(11) and (14) takes the form $%
\Gamma =\epsilon ^2NA^2(M)e(M).$ For heavy nuclei $\Gamma _A$ $<<\bar M$ and
$\bar M-M<<\bar M$. Substituting in Eq.(13) $A^2(M_X)\approx (2\pi )\delta
(M_X-\bar M)$ one obtains $e(M)\approx e(\bar M)\approx 1$. In agreement
with Eq.(3), we get $\Gamma \approx \epsilon ^2NA^2(M)$.

It is instructive to analyze the decay process in the time-dependent
S-matrix approach. In the interaction representation with respect to the
strong Hamiltonian $H_0$, the state $|N,Z>$ evolves according to the law $%
|N,Z,t>=S_I(t,0)|N,Z,0>$ where $|N,Z,0>=|N,Z>$ and $S_I(t,0)=T\exp (-i\int
V_I(t^{\prime })dt^{\prime })$. The value $V_I$ is defined by Eq.(6) with $%
U(t,0)=\exp (-iH_0t)$. To the second order in the $V$, the decay amplitude
takes the form

\begin{equation}
\label{16}<N,Z,0|N,Z,t>=1-\int_0^tdt_1%
\int_0^{t_1}dt_2<N,Z|Ve^{-i(H_0-M)(t_1-t_2)}V|N,Z>.
\end{equation}
Inserting complete set of the states $X$ to the matrix element in the right
hand side of this expression, we get

\begin{equation}
\label{17}<N,Z|Ve^{-i(H_0-M)t}V|N,Z>=\epsilon ^2N\int_{}^{}\frac{dM^{\prime
} }{2\pi }A^2(M^{\prime })e^{-i(M^{\prime }-M)t}e(M^{\prime }).
\end{equation}
Here the use is made of Eqs.(11) and (14). The values of the $M^{\prime }$
are grouped around the $\bar M$ by the Lorentz function $A^2(M^{\prime })$.
Setting $e(M^{\prime })\approx e(\bar M)\approx 1$ and performing the
integration, we obtain

\begin{equation}
\label{18}<N,Z|Ve^{-i(H_0-M)t}V|N,Z>\approx \epsilon ^2Ne^{-(\Gamma
_A/2+i(\bar M-M))t}.
\end{equation}
The decay amplitude (16) at $t>>1/\Gamma _A$ takes the form

\begin{equation}
\label{19}<N,Z,0|N,Z,t>\approx 1-\epsilon ^2Nt/\chi .
\end{equation}
The term $2Re\{\epsilon ^2N/\chi \}=\epsilon N\xi $ appearing in the
absolute square of this amplitude can be recognized as the nuclear decay
width $\Gamma $. We derived therefore Eq.(3).

The possibility of reproducing the potential model results within the
field-theoretic S-matrix approach has recently been questioned$^{11}$. We
show that the results of Ref.11 are erroneous due to a misinterpretation of
the optical theorem.

The S-matrix determined by the Hamiltonian $H=H_0+V$ can be expanded into a
sum of the unit operator and the T-matrix: $S=1-iT$. The unitarity relation $%
SS^{+}=1$ implies $2ImT=-TT^{+}$.

The complete orthogonal basis $|X>=\{|X_b>,|X_{sc}>\}$ can be constructed
from eigenstates of the Hamiltonian $H_0$. The states $|X_b>$ are the
finite-norm bound states (nuclei), while the states $|X_{sc}>$ are the
infinite-norm scattering ones. The normalization conditions are $%
<X_b|X_b>=\delta _{bb^{\prime }},$ $<X_{sc}|Y_{sc}>=\delta (X-Y)$. The
expansion of the unit takes the form

\begin{equation}
\label{20}1=\sum_b|X_b><X_b|+\int d\mu (X_{sc})|X_{sc}><X_{sc}|.
\end{equation}

The value $<X_b|TT^{+}|X_b>$with the help of Eq.(20) can be represented in
the form

\begin{equation}
\label{21}<X_b|TT^{+}|X_b>=W_b+|<X_b|(-iT)|X_b>|^2
\end{equation}
where

\begin{equation}
\label{22}W_b=\sum_{b^{\prime }\neq b}|<X_b|S|X_{b^{\prime }}>|^2+\int d\mu
(X_{sc})|<X_b|S|X_{sc}>|^2.
\end{equation}
Here the orthogonality is taken into account: $<X_b|(-iT)|X_{b^{\prime
}}>=<X_b|S|X_{b^{\prime }}>$ for $b^{\prime }\neq b$ and $%
<X_b|(-iT)|X_{sc}>=<X_b|S|X_{sc}>$. The value $W_b$ is the total decay
probability of the state $X$$_b$. The second term in Eq.(21) can be
transformed as $|<X_b|(-iT)|X_b>|^2=$ $|<X_b|S|X_b>|^2-2Re<X_b|S|X_b>+1$.
The value $|<X_b|S|X_b>|^2=1-W_b$ is the probability of finding the system t
sec after its preparation in the initial state $|X_b>$. Eq.(21) becomes

\begin{equation}
\label{23}<X_b|TT^{+}|X_b>=2(1-Re<X_b|S|X_b>).
\end{equation}
For the finite-norm states, the quantity $<X_b|TT^{+}|X_b>$ has no
probability meaning, along with the imaginary part of the T-matrix element $%
-2Im<X_b|T|X_b>$.

The value $<X_{sc}|TT^{+}|X_{sc}>$ with the help of Eq.(20) is represented by

\begin{equation}
\label{24}<X_{sc}|TT^{+}|X_{sc}>=W_{sc}
\end{equation}
where
\begin{equation}
\label{25}W_{sc}=\sum_b|<X_{sc}|S|Y_b>|^2+\int d\mu
(Y_{sc})|<X_{sc}|S|Y_{sc}>|^2.
\end{equation}
The term analogous to the second term in Eq.(21) does not occur because of
zero measure of the infinite-norm scattering states. The value $W_{sc}$ is
the total transition probability to the states $Y\neq X_{sc}$. Eq.(24) is
the optical theorem.

The excited nuclei correspond to quasistationary states. These states
represent localized wave packets and have finite norm like the bound states.
One can show that for quasistationary states the relation (23) holds true
with the substitution $|X_b>\rightarrow |Q>.$ There is a deep analogy
between properties of bound states and quasistationary states (see Ref.12,
Ch.34). The parallel can also be drawn towards absence of the probability
meaning for imaginary parts of the T-matrix diagonal amplitudes. The optical
theorem takes place for the infinite-norm scattering states only. Notice
that nuclei are bound states with respect to the strong Hamiltonian $H_0$
and quasistationary ones with respect to the total Hamiltonian $H$ including
the baryon-number-violating part.

It is worthwhile to make tree remarks. (i) The decay amplitude of a
quasistationary state is given by the S-matrix element $<Q|S|Q>$. Up to a
phase factor, $<Q|S|Q>=e^{-\Gamma _Qt/2}$, in which case $%
<Q|TT^{+}|Q>=2(1-e^{-\Gamma _Qt/2})$. In Ref.10, the value $<Q|TT^{+}|Q>$ is
misinterpreted as the decay probability and an expression $%
<Q|TT^{+}|Q>=1-e^{-\Gamma _Qt}$ is erroneously suggested. It is equivalent
to the decay law $<Q|S|Q>=(1+e^{-\Gamma _Qt})/2$ which has nothing in common
with the time behavior of the decay amplitude for quasistationary states. In
particular, at $t$ $\rightarrow \infty $ the quarter of the state $|Q>$
survives. So, conclusions following Eq.(16) of Ref.11 are erroneous. (ii) It
is claimed also that the infinite-time field-theoretic S-matrix approach is
infrared divergent. In Fig.1 of Ref.11, the neutron line of the Green's
function belongs to the nuclear Bethe-Salpeter wave function, and so the
neutron is out of the mass shell. The decay amplitude therefore is not
singular. Indeed, we showed that the infinite-time S-matrix approach
(Eqs.(11) to (15)) reproduces the result of the potential model (Eqs.(5) to
(10)) and the result of the finite-time S-matrix approach (Eqs.(16) to
(19)). (iii) The exponential decay law for quasistationary states takes
place at $1/E<t<log(E/\Gamma )/\Gamma $ where $E$ is energy release in the
process (Ref.12, Ch.30). In our case $E\approx 2$ $GeV$ and the value $%
\Gamma $ is given by Eq.(3). The time interval for the exponential decay can
be evaluated to be $10^{-25}<t<10^{38}$ $sec$ for $\tau =1$ $yr$, $\Gamma
_A=100$ $MeV$, and $N=100$. At an interval $1/E<<t<<1/\Gamma $, the decay
probability is linear in time ($e^{-\Gamma _Qt}=1-\Gamma _Qt+...$). The
quadratic time dependence$^{11}$ of the nuclear decay rate is in
contradiction with the exponential decay law for quasistationary states.

To clarify physical nature of the decay rate suppression, we wish to discuss
gedankenexperiment with deuterium ($^2H$) atoms in the spin $j=1/2$ state,
which exhibits transparent mechanisms for the spin-precession-rate
suppression.

Let us consider sequence of the Stern-Gerlach magnets (SG) which measure the
spin-$z$ projection and merge the outgoing atomic beams together (for a
description such magnets see Ref.13). The SG magnets are placed along the $x$
axis at equal distances $l_0$ one from another, so that the atoms
propagating along the straight line pass successively all SG magnets. The
atoms entering the first magnet are polarized along the $z$ axis. The
existence of the magnetic field ${\bf H}=(H_x,0,0)$ with a small component $%
H_x$ along the atomic beam (version I) is assumed outside the magnets. The
atom spin precession is described by the Hamiltonian $H=-\mu {\bf \sigma H}$
with $\mu $ being the atomic magnetic moment ($\mu \approx \mu _B/3$ where $%
\mu _B$ is the Bohr magneton). The atom has spin up at a time $t$ with
probability

\begin{equation}
\label{26}w_I(t)=\cos {}^2(\omega _xt)
\end{equation}
where $\omega _x=-\mu H_x$. The neutron oscillation is described by the same
equation with the substitution $\xi \leftrightarrow \omega _x$.

In the next experiment, the existence of the magnetic field ${\bf H}%
=(H_x,0,H_z)$ with $H_x<<H_z$ (version II) is assumed outside the magnets.
The SG magnets are supplied also by detectors for the spin-down atoms. The
time evolution of the spin-up atoms with the switched off detectors is given
by equation

\begin{equation}
\label{27}w_{II}(t)=1-\frac{\omega _x^2}{\omega ^2}\sin {}^2(\omega t)
\end{equation}
where{\bf \ }${\bf \omega }=-\mu {\bf H},\omega =|{\bf \omega }|$. If
detectors are switched on, the evolution law (27) is no longer valid. The
operating detector installed at the k-th SG magnet can produce no signal
when a deuterium atom passes through it. In such a case, we know with
certainty that the atom has its spin up. The passage of the atom is
accompanied by the wavefunction collapse. Afterwards, the atomic
wavefunction up to a normalization factor coincides with the initial one,
and the oscillation process starts from the beginning. The probability that
the first N detectors all give no signals is given by

\begin{equation}
\label{28}w_{II}(t)_{on}=(1-\frac{\omega _x^2}{\omega ^2}\sin {}^2(\omega
t))^N
\end{equation}
where $N=t/t_0>>1$, $t_0=l_0/v$ is the time needed to pass from one magnet
to the next, and $v$ is the atom velocity. The time evolution of atoms in
between the SG magnets is described quantum mechanically, whereas at $t>>t_0$
due to the measurement procedure, the classical rules should be applied for
calculation of the probability. Each instant of time $t=kt_0$ the atom is
either detected with a probability $1-w_{II}(t_0)$ or continue its
propagation with a probability $w_{II}(t_0)$. To derive Eq.(28), one should
take $N$ times the product of the elementary probabilities $w_{II}(t_0)$.
The value $w_{II}(t_0)$ is close to unity, since $\omega _x<<\omega $, so
the probability to survive in the spin-up state follows the exponential
decay law

\begin{equation}
\label{29}w_{II}(t)_{on}\approx \exp (-\omega _x\xi ^{\prime }t)
\end{equation}
where $\omega _x>0$ is assumed and

\begin{equation}
\label{30}\xi ^{\prime }=\omega _xt_0\frac{\sin {}^2(\omega _zt_0)}{(\omega
_zt_0)^2}
\end{equation}
to the first order in the $\omega _x$.

In the first experiment, the value $\omega _x$ determines precession rate of
the spin-up atoms. In the second experiment, new dimensionless parameter $%
\xi ^{\prime }<<1$ occurs. The decay rate for the spin-up atoms turns out to
be $\omega _x\xi ^{\prime }<<\omega _x$.

The apparent analogy between the neutron oscillation and the atomic spin
precession is summarized below.

\vspace{0.5cm} $
\begin{array}{ccc}
\mbox{neutrons} & \longleftrightarrow & \mbox{spin-up atoms} \\
\mbox{antineutrons} & \longleftrightarrow & \mbox{spin-down atoms} \\
\mbox{vacuum
n-\=n oscillation} & \longleftrightarrow & \mbox{atom spin precession (v.I)}
\\ \mbox{vacuum oscillation rate }\epsilon \ \mbox{of neutrons} &
\longleftrightarrow & \mbox{spin precession rate }\omega _x%
\mbox{ of atoms
(v. I)} \\ \Delta U=\bar M-M & \longleftrightarrow & \omega _z\
\mbox{(v.\ II)} \\ \mbox{\=n length free path in nuclear matter} &
\longleftrightarrow & \mbox{distance }l_0\mbox{ between SG magnets (v.II)}
\\ \mbox{antineutron annihilation} & \longleftrightarrow &
\mbox{detecting
spin-down atoms (v.II)} \\ \mbox{suppression parameter }\xi &
\longleftrightarrow & \mbox{suppression parameter }\xi ^{\prime }\
\mbox{(v.II)} \\ \mbox{nuclear decay rate per neutron }\epsilon \xi &
\longleftrightarrow & \mbox{atom spin-up decay rate }\omega _x\xi ^{\prime }%
\mbox{ (v.II)}
\end{array}
$

\vspace{0.5cm}

Reasons, according to which physical results in these two problems are
identical, can be explained by considering the atomic spin dynamics
described by the Hamiltonian (5) with the time dependent width

\begin{equation}
\label{31}\Gamma (t)=\sum_k\alpha \delta (t-kt_0).
\end{equation}
Here the k's are integer numbers. In the intervals $kt_0<t<(k+1)t_0$ the
atom is out of the SG magnets and its spin precession is governed by the
magnetic field ${\bf H}=(H_x,0,H_z)$. At the time $t=kt_0$ the act of the
measurement takes place and the spin-down component of the wavefunction
vanishes. In order to set the lower component equal to zero, we must pass to
the limit $\alpha \rightarrow \infty $. The upper component at $t=kt_0$
remains continuous. The probability to find the deuterium atom in the
spin-up state can be computed according to the usual rule as the modulus
squared of the upper component of the atomic wavefunction. Using such a
prescription, the results of Eqs.(28) and (29) can easily be reproduced.

One can show that the S-matrix projection $S(t,0)P_{+}$ to the spin-up
initial state, being averaged over an interval $(t-T,t+T)$ at $T>>t_0$,

\begin{equation}
\label{32}\overline{S(t,0)}P_{+}=\frac 1{2T}\int_{-T}^Tdt^{\prime
}S(t^{\prime },0)
\end{equation}
coincides with the S-matrix projection for the constant width $\Gamma _A$.
It means that projection $\overline{S(t,0)}P_{+}$ responsible for
disappearance of atoms from the beam can be reproduced in a model identical
to the potential model for description of the n-\=n oscillation in nuclei.

In the limit $\alpha \rightarrow \infty $, the functions entering Eq.(8) can
easily be computed. After averaging over the time, we get the following
expression

\begin{equation}
\label{33}\overline{S(t,0)}P_{+}=e^{i\omega _zt}(P_{+}-i\omega _x/\chi
^{\prime }\sigma _{-}-\omega _x^2t/\chi ^{\prime }P_{+}+...)
\end{equation}
where $1/\chi ^{\prime }$ $=(sin^2(\omega _zt_0)-i((\omega _zt_0)-sin(\omega
_zt_0)cos(\omega _zt_0)))/(2\omega _z^2t_0)$. Notice that $\omega _xRe$ $%
\{2/\chi ^{\prime }\}=\xi ^{\prime }$. Eq.(33) is identical to Eq.(10) for $%
\epsilon =$ $\omega _x$ and $\chi =\chi ^{\prime }$. Given that $\Gamma _A$
and $\Delta U$ are known, one can find $t_0$ and $\omega _z$. To the second
order in the $\omega _x$ ($\epsilon $), the evolution laws in these two
problems are identical in the average sense.

The qualitative explanation for the neutron-decay-rate suppression can be
given in the following way. In the case $\Delta U>>\Gamma _A$ one can start
from Eq.(31) that gives average admixture $\approx \omega _x^2/\omega _z^2$
of the spin-down atoms in the beam. The rate of disappearance of the atoms
can be estimated to be $\approx \omega _x^2/(\omega _z^2t_0)<<\omega _x$.
The equation $\chi =\chi ^{\prime }$ gives $\omega _z=\Delta U/2$, $%
t_0\approx 1/\Gamma _A>>2/\Delta U$. Using the above analogy, we derive the
corresponding neutron decay rate $\approx \epsilon ^2\Gamma _A/\Delta
U^2=\epsilon \xi <<\epsilon $.

The inequality $\Delta U\neq 0$ is not unique reason for suppression of the
decay rate. Let $\Delta U=0$. The equation $\chi =\chi ^{\prime }$ gives $%
\omega _z=0$ and $t_0=4/\Gamma _A$. The decay rate of the spin-up atoms
becomes $\omega _x\xi ^{\prime }=\omega _x^2t_0$ $<$$<$ $\omega _x$. The
detection of the spin-down atoms suppresses the spin precession process by
itself. The neutron propagation in nuclei is accompanied by interactions
with surrounding nucleons. The meson exchanges take place each $t_0=4/\Gamma
_A\approx 10^{-24}$ $sec$. They can be interpreted as acts of measurements
of the lower antineutron component of the wavefunction (2). These
interactions are followed by collapse of the wavefunction to the pure
neutron state. The wavefunction collapse in turn reduces the nuclear decay
rate down to $4\epsilon ^2/\Gamma _A=\epsilon \xi <<\epsilon $.

The analogous suppression mechanism exists for ultracold neutrons in a trap.
The absorption probability for antineutrons in collisions with the copper
boundary is estimat-ed$^3$ to be $\approx 1/5$ for the tangent velocities $%
\approx 4m/sec$. The suppression of the precession rate occurs due to the
wavefunction collapse that takes place with certainty under those conditions
one time per $\approx 5$ neutron collisions with the trap boundary. The
suppression factor is about $5\epsilon t_0<10^{-6}$, where $t_0$ ($\approx
seconds$) is a typical time between two collisions.

In the conclusion, the field-theoretic S-matrix approach allows to justify
the results of the potential model for heavy nuclei. The suppression of the
nuclear decay rate is the result of the energy gap $\Delta U$ and the
antineutron annihilation width $\Gamma _A$. The suppression of the neutron
oscillation process is not specific for nuclear physics. In the SG-type
experiment described above, the atom spin precession is suppressed due to
the similar physical reasons. For light nuclei, the procedure$^{4,5}$ of
solving the inhomogeneous Schr\"odinger equation for antineutron
wavefunction in the optical potential is more adequate then the direct use
of Eq.(3). Being applied for heavy nuclei, such a procedure reproduces
Eq.(3). The models$^{1-7}$ account to a various degree of accuracy for
dynamics of the antineutron creation and annihilation in nuclei. Further
refinement of parameters of the antineutron optical potential entering
Eq.(3) would be desirable. It is unlikely, however, that the corresponding
numerical estimates are changed more then by an order of magnitude.

After submitting this work our attention has been called to a recent paper
[14] in which results of Ref.11 obtained with the use of the field-theoretic
methods are critisized on the basis the optical potential model. The authors
give two new derivations for expression for the nuclear decay width within
the optical potential model (Eqs.(1) to (5) and Eqs.(11) to (14)). We argued
that the imaginary part of the T-matrix has no probability meaning for bound
and quasistationary states. It is not clear then why the value $W_n$ in
Eq.(3) of Ref.14 is not a probability whereas the value $W$ in Eq.(5) does.
These values both are defined, respectively, as imaginary parts of strong
and baryon-number violating (GUT's) T-matix amplitudes. We discussed the
T-matrix properties for Hermitian Hamiltonians. The optical model
Hamiltonian $H=-i\Gamma /2$ used in Ref.14 is not Hermitian, and so it is
necessary to reanalyze the problem on slightly more general grounds.

Notice that Eq.(16) of Ref.11 which represents a finite-time analog of
Eq.(4) of Ref.3 is correct and it remains valid for non-Hermitian
Hamiltonians also, since hermiticity of the Hamiltonian is not assumed in
deriving this equation. It is allowed therefore to substitute here $%
H=-i\Gamma /2$. In this way, purely imaginary T-matrix occurs: $T=-iW/2$
with $W=\epsilon \xi t$ for $t>>1/\Gamma _A$. The S-matrix becomes $%
S=1-iT=1-W/2$. The probability of finding the system in the initial state $t$
sec after its preparation becomes $|S|^2=1+2Re(-iT)$ $+TT^{+}$ $\approx
1+2ImT=$ $1-W$. The quantity $W=-2ImT$ thus really receives, to the first
order in $T=O(t)$, a meaning of the decay probability. We can expand the
value $W_n$ in Eq.(3) of Ref.14 in power series in $t(=t_\alpha -t_\beta )$
and verify, in agreement with that conclusion, that the lowest order term
gives the decay probability $\Gamma _{A\ }t<<1$. The reason stems from the
fact of neglecting the small term $TT^{+}=O(t^2)$ in the expression for the $%
|S|^2$. At high times, however, $W_n\rightarrow 2$. This is no longer
surprise, since $TT^{+}$ becomes high as $t$ increases, and so it should be
taken into account in the $|S|^2$. The value $W$ defined by Eq.(1) in Ref.14
is calculated to the first order in $t$. This is a reason for it gives the
correct result for the probability. The next order term in $t$, however,
should already be wrong, if the value $TT^{+}$ is disregarded. The
instability of quasistationary states implies $<Q|S|Q>\rightarrow 0$ when $t$
increases. Our Eq.(23) gives then $W_n\rightarrow 2$ at $t>>1/\Gamma _A$ and
$W\rightarrow 2$ at $t>>1/\Gamma $.

Therefore, the value $W_n$ in Eq.(3) of Ref.14 is not a probability, when it
is computed to all orders in $t$. To the first order in $t$ it acquires the
probability meaning. The value $W$ in Eq.(5) is not a probability also. It
coincides with the probability to the first order in $t$ only.

\vspace{0.5cm}

The author is indebted to K.G.Boreskov, B.V.Martemyanov and M.G.Schepkin for
useful discussions, A.Gal for communication on the paper [14], and
especially to L.A.Kondratyuk for attracting the interest to the problem of
the neutron oscillations and valuable discussions of the optical potential
model. This work is supported by the Russian Fund for Fundamental Researches
under the Grant No. 94-02-03068.

\end{document}